\documentstyle[aps,prb,epsf,epsfig,graphics]{revtex}

\newcommand \be {\begin{equation}}
\newcommand \ee {\end{equation}}
\newcommand \bea {\begin{eqnarray}}
\newcommand \eea {\end{eqnarray}}
\newcommand \eps {\epsilon}
\newcommand \bi {\bibitem}
\newcommand \s {\sigma}

\newcommand \dd {{\mathrm{d}}}

\begin{document}

\title{Chaotic, memory and cooling rate effects in spin glasses: Is
the Edwards-Anderson model a good spin glass?}
 
\author{Marco Picco\protect\cite{marco}}
\address{
	LPTHE\\
	Universit\'e Pierre et Marie Curie, Paris VI\\
	Universit\'e Denis Diderot, Paris VII\\
	Boite 126, Tour 16, 1$^{\it er}$ \'etage, 4 place Jussieu\\
	F-75252 Paris Cedex 05, France}

\author{Federico Ricci-Tersenghi\protect\cite{federico}}
\address{
	Abdus Salam International Center for Theoretical Physics,
	Condensed Matter Group\\
	Strada Costiera 11, P.O. Box 586, I-34100 Trieste, Italy}

\author{Felix Ritort\protect\cite{felix}}
\address{
	Department of Physics, Faculty of Physics,
	University of Barcelona\\
	Diagonal 647, 08028 Barcelona, Spain}

\date{\today}

\maketitle

\begin{abstract}
We investigate chaotic, memory and cooling rate effects in the three
dimensional Edwards-Anderson model by doing thermoremanent (TRM) and
AC susceptibility numerical experiments and making a detailed
comparison with laboratory experiments on spin glasses.  In contrast
to the experiments, the Edwards-Anderson model does not show any trace
of re-initialization processes in temperature change experiments (TRM
or AC).  A detailed comparison with AC relaxation experiments in the
presence of DC magnetic field or coupling distribution perturbations
reveals that the absence of chaotic effects in the Edwards-Anderson
model is a consequence of the presence of strong cooling rate effects.
We discuss possible solutions to this discrepancy, in particular the
smallness of the time scales reached in numerical experiments, but we
also question the validity of the Edwards-Anderson model to reproduce
the experimental results.
\end{abstract}


\section{Introduction}

One of the most characteristic effects in disordered or glassy systems
in their non-stationary regime is the presence of aging. The response
of the system stiffens with age showing that it depends on all the
previous history through e.g.\ the waiting time~\cite{POLYMERS,LSNB}.
Experimentally, this phenomenon is well documented through
magnetization relaxation experiments and AC susceptibility
measurements~\cite{SITGES,NS,BCKM}.  Despite of the different
experimental procedures needed for both type of measurements,
magnetization relaxation and AC susceptibility give similar
information regarding the aging behavior and its waiting time
dependence.

On top of all these non-equilibrium phenomenology, recent dynamical
experiments in spin glasses show very peculiar chaotic (also called
rejuvenation), memory as well as cooling rate
effects~\cite{MEMORY_EXP,RECENT_SG,JJJN}. These effects are thought to
be the signature of the spin glass state being much different to those
found in usual ferromagnets or other disordered
systems~\cite{MNF}. The most unusual experimental result in spin
glasses is the absence of cooling rate effects. The approach to
equilibrium at a given temperature after cooling from high
temperatures is not influenced by the whole cooling history at higher
temperatures but only by the time spent at the last temperature in the
thermal history.  Experimentally, rejuvenation or chaotic effects in
spin glasses are measured in a clear way by doing AC measurements. An
alternating magnetic field of frequency $\omega=\frac{1}{P}$ where $P$
is the period is applied to the sample and both components of the AC
susceptibility (the {\em in-phase} $\chi'$ and the {\em out-of-phase}
$\chi''$) are measured.

Although the major part of these measurements have been done on
insulating spin glasses they are common also to metallic spin glasses
leading to the question whether these effects are also present in the
most well known theoretical models. Despite of the large amount of
theoretical work devoted to aging effects in glasses and spin glasses
there is still no convincing and final explanation for the origin of
these peculiar chaotic and memory effects.  The comprehension of these
effects will certainly provide a clue to the understanding of the
nature of the glassy state.

Because memory and chaotic (or rejuvenation) effects are intrinsic to
spin glasses (metallic and insulating) it is important to understand
whether models for spin glasses are able to reproduce the experimental
results. It is widely accepted that the Edwards-Anderson model
contains the main features observed in real spin glasses. The purpose
of this paper is to present a detailed and critical study of these
phenomena in the Edwards-Anderson model in three dimensions. This is
not a simple matter to address. Despite of the large amount of
numerical studies on equilibrium and non-equilibrium phenomena there
is no clear evidence that the Edwards-Anderson model reproduces the
main results found in experiments. Note that, even the question
whether there is or not phase transition in the 3D Edwards-Anderson
model is still not fully settled~\cite{MPR,PY}.

The purpose of this paper is to present a numerical investigation,
fully experimentally oriented, of the non-equilibrium behavior of the
three dimensional Ising spin glass with special emphasis on recent
experiments where memory and chaos effects where found.  This question
is of the out-most importance concerning modeling. If some
experimentally observed results are missing in any theory then we must
understand why. There have been several investigations in the
literature devoted to this subject, but still a clear answer is
missing~\cite{KSSR,RIEGER,KYT}. Here we will provide a complementary
investigation to results already published, emphasizing the
experimental results and comparing different types of experiments.  In
particular, our main effort will be devoted to investigate
thermoremanent and AC numerical experiments. Although thermoremanent
studies have been largely considered in the past there are very few
numerical investigations devoted to the AC topic~\cite{SUECS_NUMERIC}.

The paper is divided as follows. In Section~\ref{sec:model} we
introduce the model as well as the dynamical
procedure. Section~\ref{sec:obs} discusses the two type of
measurements we have done: magnetization relaxation and AC numerical
experiments. Sections~\ref{sec:chaos_TRM} and~\ref{sec:chaos_AC}
present a detailed investigation of memory and chaotic effects with
thermoremanent and AC experiments respectively. Finally we present a
discussion of the results.

\section{The Edwards-Anderson model and some details of the
simulation}
\label{sec:model}

The Edwards-Anderson model~\cite{EA} was proposed in the early 70's as
the simplest model which contains the main ingredients relevant to
explain the spin-glass phenomenology. In particular, it displays a
phase transition characterized by the onset of freezing in spin-spin
correlations and a divergent non-linear
susceptibility~\cite{REVIEW_SG}. The model is defined by the following
Hamiltonian
\begin{equation}
{\cal H}=-\sum_{(i,j)}\,J_{ij}\s_i\s_j -h\sum_{i=1}^V\s_i \quad ,
\label{eqEA}
\end{equation}
where the indices $i,j$ run from 1 to V, the $\s_i$ are Ising spins
and the pairs $(i,j)$ identify nearest neighbors in a finite
dimensional lattice. The exchange couplings $J_{ij}$ are taken from a
random distribution. To avoid degeneracy of the ground state the
simplest choice is a Gaussian distribution with zero average and
finite variance,
\begin{equation}
{\cal P}(J) = \bigl( \frac{1}{2\pi\Delta^2} \bigr)^{\frac{1}{2}} \,
\exp\bigl(-\frac{J^2}{2\Delta^2}\bigr) \quad .
\label{eqP}
\end{equation}

The model is defined in any number of finite dimensions although our
main concern here is the three-dimensional case where there is a spin
glass transition at finite temperature $T_c\simeq 0.95
\Delta$~\cite{MPR}.  Hereafter, unless differently specified, we will
consider $\Delta=1$ without loss of generality.

Monte Carlo simulations of (\ref{eqEA}) use random updating of the
spins with the Metropolis algorithm. A spin is randomly chosen and its
value changed with the proper probability.
Dynamical experiments use very large lattices (typical sizes are in
the range $L=20-100$) with negligible finite-size effects for the
largest sizes ($L=64$ for magnetization relaxation experiments and
$L=100$ for AC experiments).  Here we present two classes of different
but related experiments. Magnetization and correlation relaxation
simulations have run on a special purpose machine APE~\cite{APE} for
sizes $64^3$ and averaging over 10 or 100 samples. AC experiments were
run for a single sample on a Linux cluster of PC's for sizes
$L=64,100$.

Before presenting the results it is convenient to discuss the fidelity
of the EA model to real spin glasses. Clearly, the EA model is an
idealization of the real microscopic interaction found in spin
glasses~\cite{REVIEW_SG}.  Spin glasses are commonly distinguished
into two large classes: metallic and insulating. Metallic spin glasses
are diluted magnets where a metallic host matrix is doped with some
ferromagnetic impurities (for instance AgMn, AuFe, CuMn). In these
systems spin interactions are due to indirect exchange and mediated
through conduction electrons (the RKKY interaction). Metallic spin
glasses are then diluted magnets where site disorder induces
frustrated short-ranged interactions (decaying like $1/r^3$ with $r$
being the distance between impurities). Insulating spin glasses are
much different. In this case, exchange interactions are usually
antiferromagnetic between neighbor spins but dilution and defects lead
to a strong frustration. Apart from the different microscopic origin
of the frustrating interaction, spins are really Heisenberg-like and
the Ising behavior arises from the uniaxial anisotropy present in
these type of systems. Because anisotropy is usually strong and the
local rotational symmetry of Heisenberg spins is broken, a treatment
taking pure Ising spins turns out to be a good approximation \cite{HEIS}.
Having in mind these limitations, the Hamiltonian (\ref{eqEA}) is the
simplest model which contains disorder and frustration, the two
ingredients commonly found in real spin glasses.

\section{Magnetization relaxation and AC experiments}
\label{sec:obs}

There are two alternative but equivalent ways to experimentally
investigate non-equilibrium phenomena in spin glasses: magnetization
relaxation experiments and AC measurements. A very complete
description of these methods can be found in~\cite{SITGES}. Here we
only remind the main results.

\begin{itemize}

\item{Magnetization relaxation (TRM) experiments}

Relaxation measurements are done applying a uniform magnetic field and
measuring the decay of the thermoremanent magnetization (hereafter
referred to as TRM), equivalently, the growth of the zero-field cooled
magnetization. The typical experiment consists in the following. A
sample is fastly quenched below the spin glass transition temperature
for a time $t_w$ (i.e.\ the waiting time). Then a uniform small
magnetic field $h$ is applied and the growth of the magnetization
measured,
\begin{equation}
\chi(t_w,t_w+t)=\frac{1}{Vh}\sum_{i=1}^V\s_i(t_w+t) \quad .
\label{eqM}
\end{equation}
In the linear response regime Eq.(\ref{eqM}) can be written as
\begin{equation}
\chi(t_w,t_w+t)=\int_{t_w}^{t_w+t}R(t_w+t,s) \dd s \quad ,
\label{eqR}
\end{equation}
where $R(t,s)$ is the response function which gives the change of the
magnetization $\delta M$ at time $t$ when a pulse of the magnetic
field $\delta h$ is applied at previous time $s$.  In spin glasses
aging manifests by the fact that $\chi(t_w,t_w+t)$ shows a strong
dependence on the value of $t_w$. In general, one finds the following decomposition
\begin{equation}
\chi(t_w,t_w+t)=\chi_{st}(t)+\chi_{ag}(t_w,t_w+t) \quad ,
\label{eqstag}
\end{equation}
where $\chi_{st}$ and $\chi_{ag}$ are respectively the stationary and
aging parts. Experimentally the aging part approximately scales with
the waiting time $t_w$ in the following way
\begin{equation}
\chi_{ag}(t_w,t_w+t)=f(t/t_w) \quad ,
\label{eqmag}
\end{equation}
although systematic deviations from this scaling behavior have been
observed. This point will be discussed later on.

Related to magnetization another quantity of interest which can be
numerically investigated in simulation are two-time
correlations. These quantities are difficult to experimentally measure
in spin glasses but very easy to compute in simulations. They are
defined by
\begin{equation}
C(t_w,t_w+t)=\frac{1}{V}\sum_{i=1}^V\s_i(t_w)\s_i(t_w+t) \quad .
\label{eqC}
\end{equation}
Again, in the non-equilibrium regime Eq.(\ref{eqC}) can be decomposed
in two pieces, a stationary part plus an aging part
\begin{equation}
C(t_w,t_w+t)=C_{st}(t)+C_{ag}(t_w,t_w+t) \quad .
\label{eqcstag}
\end{equation}
Similarly to the magnetization, the aging part of the correlation
is approximately described by the following scaling behavior
\begin{equation}
C_{ag}(t_w,t_w+t)=g(t/t_w) \quad ,
\label{eqcag}
\end{equation}
again with systematic (but small) deviations respect to it. The
stationary part of the correlation and magnetization are related
through the fluctuation-dissipation theorem (FDT)
\begin{equation}
\chi_{st}(t)=\frac{1-C_{st}(t)}{T} \quad .
\label{FDT}
\end{equation}
Although in mean-field spin glasses a more general relation seems to
be valid~\cite{BCKM}. It links response and correlation functions also
in the off-equilibrium regime through
\begin{equation}
X[C] = -T \left. \frac{\partial\chi(s,t)}{\partial C(s,t)}
\right|_{C(s,t)=C(t_w,t_w+t)} \quad ,
\label{GFDT}
\end{equation}
where the fluctuation-dissipation ratio $X$ depends only on the
correlation function in the large times limit ($t,t_w\to\infty$).  In
the quasi-equilibrium regime ($t<t_w$) we have that $X=1$ and we
recover the usual FDT. In the aging regime ($t>t_w$) the ratio is
smaller than one $X<1$ and it can be interpreted as a larger effective
temperature $T_{\rm eff} \sim T/X$.  In finite-dimensional spin
glasses the validity of Eq.(\ref{GFDT}) has been numerically
checked~\cite{FR,MPRR} and it has been related to the equilibrium
distribution of overlaps~\cite{MPRR}.

\item{AC measurements}

In these experiments an oscillating magnetic field
$h(t)=h_0\cos(2\pi\omega t)$ of frequency $\omega=\frac{1}{P}$, where
$P$ is the period, is applied to the system and the magnetization
measured as a function of time
\begin{equation}
M(t)=M_0\cos(2\pi\omega t+\phi) \quad ,
\label{eqMAC}
\end{equation}
where $M_0$ is the intensity of the magnetization and $\phi$ is the
dephasing between the magnetization and the field. The origin of the
dephasing is dissipation in the system which prevents the
magnetization to follow the oscillations of the magnetic field. From
the magnetization one can obtain the in-phase and out-of-phase
susceptibilities defined as
\begin{eqnarray}
\chi'=\frac{M_0\cos(\phi)}{h_0}=\frac{2\int_0^PM(t)\cos(2\pi\omega
t)dt}{h_0} \quad , \label{eqchi1}\\
\chi''=\frac{M_0\sin(\phi)}{h_0}=\frac{2\int_0^PM(t)\sin(2\pi\omega
t)dt}{h_0} \quad . \label{eqchi2}
\end{eqnarray}
The dephasing $\phi$ measures the rate of dissipation in the system
and is given by
\begin{equation}
\tan(\phi)=\frac{\chi''}{\chi'} \quad .
\label{tanfi}
\end{equation}

In numerical simulations the in-phase and out-of-phase
susceptibilities are computed by averaging the right-hand side in
Eqs.(\ref{eqchi1}),(\ref{eqchi2}) over several periods
$P=\frac{1}{\omega}$. This means a very large measurement time for low
frequencies for both experiments and simulations.  In the
non-equilibrium regime the AC susceptibility~\cite{note1} depends on
both the waiting time and the frequency. On general grounds one
expects that
\begin{equation}
\chi(\omega,t)=\chi_{st}(\omega)+\chi_{ag}(\omega,t) \quad ,
\label{eqchit}
\end{equation}
where the aging part of the AC susceptibility approximately satisfies
a scaling behavior
\begin{equation}
\chi_{ag}(\omega,t)\sim h(\omega t) \quad .
\label{eqchiag}
\end{equation}

\end{itemize}

Both types of measurements give equivalent information about the
relaxation dynamics but in different time sectors. As discussed in
ref.~\cite{SITGES}, TRM experiments give information on time scales
ranging between the two limits $t \ll t_w$ and $t \gg t_w$. For AC
experiments the frequency $\omega$ corresponds to the inverse of the
observation time $t_{obs}$ (note that in TRM experiments after
switching the field we have that $t_{obs}=t$) while the age $t_a$
corresponds to the total elapsed time $t+t_w$. In AC experiments in
order to get reliable results on $\chi'$ and $\chi''$ one needs to
average over several periods of the field, while keeping the age of
the system more or less unaltered (otherwise the two limiting regimes
would mixed and the results would be unclear). This is possible only
if $\omega t_a \gg 1$, which imply $t \ll t_w$.  Consequently, in AC
measurements one is able only to explore the beginning of the aging
regime, also called quasi-stationary regime~\cite{HOV} which
represents a smaller time window than in TRM experiments.

\section{Memory and chaos in correlation and response functions}
\label{sec:chaos_TRM}

In this Section we perform a study of memory and chaos effect in spin
glasses measuring correlation and response functions.  We always
take the measurements from 10 samples of a $64^3$ system (unless
differently specified). We closely follow the experimental procedure
on what concerns temperature changes and we keep the ratio between
time scales entering in the simulation similar to the experimental
ones (with the same limitations of sizes, magnetic fields and absolute
time scales as already discussed).

\subsection{``Cooling and stop'' experiment}

\begin{figure}
\begin{center}
\epsfxsize=0.7\textwidth
\epsffile{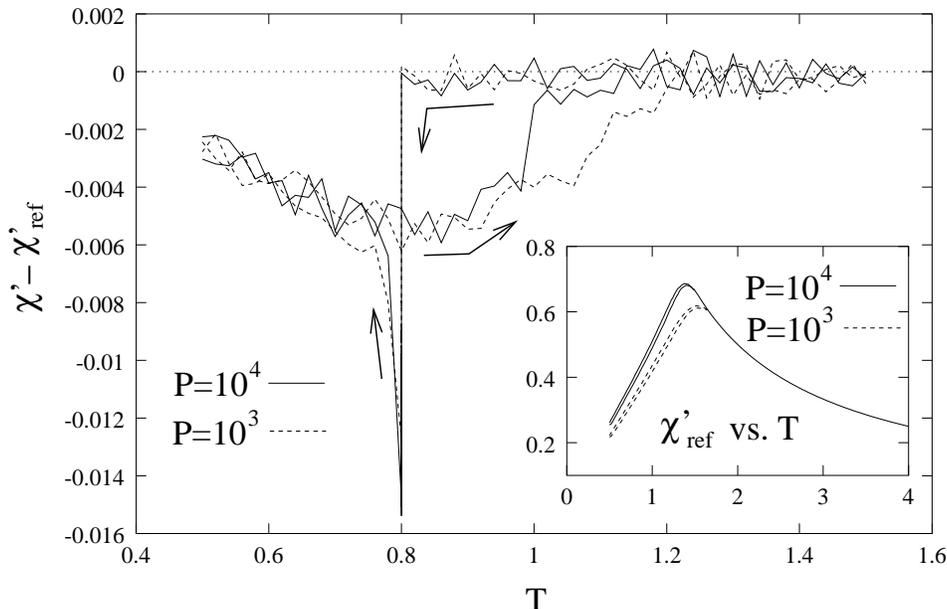}
\caption{The ``cooling and stop'' experiment in the EA model gives no
evidence for such strong chaos and memory effects experimentally found
on real samples. Here we use 100 samples of a $64^3$ system, cooling
rates inversely proportional to $P$ and probing time scales ${\cal
O}(P)$. In the inset we show the reference curves measured without any
stop during the cooling.}
\label{cool_stop}
\end{center}
\end{figure}

The experiment is performed in the following way (for more details the
reader is address to the original paper~\cite{MEMORY_EXP}).  Starting
from the high temperature phase, a spin glass sample is cooled at a
fixed cooling rate into the glassy phase. When a temperature $T^*
\simeq 0.8 T_c$ is reached, the cooling process is stopped and the
sample is let to relax for a long time. This relaxation produces a
decrease in the susceptibility (in both $\chi'$ and $\chi''$) with
respect to the reference curve (obtained with the same constant
cooling rate and without any stop).  After that long time the cooling
process is continued down to a low temperature and finally the sample
is heated back again at a constant heating rate (equal to the cooling
one) and {\em without} any stop.

There are two relevant results in this experiment. First, as soon as
the cooling process is started again after the stop, the
susceptibility merges rapidly with the reference curve, quickly
``forgetting'' the thermalization work done near the temperature $T^*$
({\em chaos} effect).  Second, when the sample is heated back at a
fixed heating rate and without any stop, the susceptibility closely
follows the cooling curve and it goes through the dip at $T^*$ ({\em
memory} effect).

In our simulations we do an analogous experiment, measuring
correlation functions instead of susceptibilities.  We divide the time
of the experiment in intervals of duration $P$ (here two values
$P=10^3,10^4$ will be considered) and we measure the correlation
function between the first and the last times in the interval,
$C(0,P)$. This correlation is strictly related to the in-phase
susceptibility, $\chi'=(1-C)/T$, measured with an external magnetic
field of frequency $\omega \propto 1/P$.  In order to confirm this
relation we show in the inset of Fig.~\ref{cool_stop} the
susceptibilities measured in two reference experiments without any
stop and with cooling and heating rates such that we perform $P$ MCS
at every temperature and then we change the temperature by $\Delta T =
0.02$. The curves resemble very much the experimental ones. Compared
to the AC measurements, simulations with the correlations have the
advantage that one can reach higher values of $P$, i.e. lower
frequencies. Note that, because of the precision required in this
experiment, all the susceptibilities have been averaged over 100
samples of size $64^3$. In the inset of Fig.~\ref{cool_stop} some
effects due to the finite cooling rates can also be appreciated. In
particular, one can see that the curves measured during the cooling
processes stay a little bit above the corresponding curves measured
during the heating process. This means that the system ``accumulates''
part of the relaxation work done in the low-temperature phase. The
cooling and heating curves merge together only when the system comes
back to the high-temperature phase. This phenomenon is present also in
experimental spin glasses, even if with a much smaller
intensity~\cite{ERIC}.

The interesting information can be obtained once we perform a long
stop in the spin-glass phase. Here we stop the cooling at a
temperature $T^*=0.8$ for $t_w=100 \cdot P$ MCS (i.e.\ 100 intervals)
and measure $\chi'$ from $C(t_w,t_w+P)$.  The system relaxes and
$\chi'-\chi^\prime_{\rm ref}$ becomes negative.  However, when we
start again cooling the system, the susceptibility does not recover
completely the reference curve and it always remain with smaller
values. Note that at the experimental level~\cite{MEMORY_EXP,JJJN} the
susceptibility recovers the reference value very rapidly (around $T
\simeq 0.65 \div 0.7$ on the scale of Fig.~\ref{cool_stop}) and that
the apparent rapid increase just below $T^*$ in Fig.~\ref{cool_stop}
is due to the very zoomed $y$ axis and it has in fact a slope of order
one. Moreover the curve followed by the data for $T<T^*$ does not seem
to depend on $P$ and if we consider the relative difference,
$(\chi'-\chi^\prime_{\rm ref})/\chi^\prime_{\rm ref}$, we would obtain
that the convergence towards 0 is still slower, due to the fact that
both susceptibilities are decreasing with $T$.

During the heating process the system stay on the same cooling curve
and it does not show any strange effect near $T^*$.  For $T>T^*$ it
finally recovers the reference curve and here is where we observe the
largest dependence on $P$. The temperature where $\chi'$ becomes
comparable with $\chi^\prime_{\rm ref}$ strongly decreases with
increasing $P$. Nevertheless for the times we have access to
($P=10^3,10^4$) this temperature is larger than the critical one $T_c
\simeq 0.95$ and in the limit of large times, $P \to \infty$, it can
converge to both $T_c$ or $T^*$.

In conclusion we can assert that the three-dimensional
Edwards-Anderson model does not show, on the time scales we have
access to, the strong memory and chaos effects real spin glasses show.

The same numerical experiment we have presented in this Section have
been recently done also by Komori {\it et al.} (see Figure 7 in
ref.~\cite{KYT}).  In that figure it is shown that some kind of memory
and chaotic effects are found when measuring the out-of-phase
component of the AC susceptibility.  But that figure and all the
subsequent authors discussion based on it are inconclusive for the
following reasons.  They use a magnetic field that oscillates too fast
($P=160$). As a consequence the effective critical temperature is very
high, $T_c^{\rm eff} > 2.5 \simeq 3\,T_c$.  Moreover they use a cooling
rate which is 10 times higher compared to experimental protocols. Each
point in their figure is a measure over a single cycle of the field
and a single dynamical history averaged over many different
samples~\cite{TAKAYAMA}.  Finally they claim to see in their figure 7
a merge of the susceptibility data to the reference curve, which is
far from evident without any zoom of the interesting region. Note that
both susceptibilities ($\chi$ and $\chi_{\rm ref}$) goes to zero when
$T \to 0$ and then we also expect $\chi-\chi_{\rm ref}$ to become
zero. One should check that the relative difference is going to zero
faster in order to claim for the presence of chaotic effects.  In our
study we had to increase the precision of more than 2 orders of
magnitude (note the $y$-axis scale in Fig.~\ref{cool_stop}) in order
to discern the effect.  The deceiving result is that, if the
experiment corresponding to figure 7 of~\cite{KYT} is done with slower
cooling rates (unfortunately, such results were not shown
in~\cite{KYT}) , one does not probably observe any trace of
rejuvenation or memory due to the strong cooling rate effects. This is
definitely different from what experiments show.

\subsection{Temperature cycling experiments}

In another set of very interesting experiments the temperature is
changed according to the following scheduling: $t_{w1}$ seconds at
$T_1$, then $t_{w2}$ at $T_2$ and finally $t_{w3}$ at $T_1$ again.
After that the TRM decay is measured.  Depending on the sign of
$\Delta T=T_2-T_1$ the system responds in different ways. For $\Delta
T < 0$ an effective waiting time, $t_w^{\rm eff}$, in the TRM decay can be
defined and it is a monotonic function of $\Delta T$ such that
$t_w^{\rm eff}=t_{w1}+t_{w2}+t_{w3}$ for $\Delta T=0$ and
$t_w^{\rm eff}=t_{w1}+t_{w3}$ for $|\Delta T|$ large.  For $\Delta T > 0$
the TRM decay follow a more complicated law and it can not be
described just by an effective waiting time.  Nevertheless is always
possible to define a correlation time, which turns out to be a
monotonic function of $\Delta T$, taking the same value as before for
$\Delta T=0$, but converging to $t_w^{\rm eff}=t_{w3}$ for large $|\Delta
T|$.

\begin{figure}
\begin{center}
\epsfxsize=0.7\textwidth
\epsffile{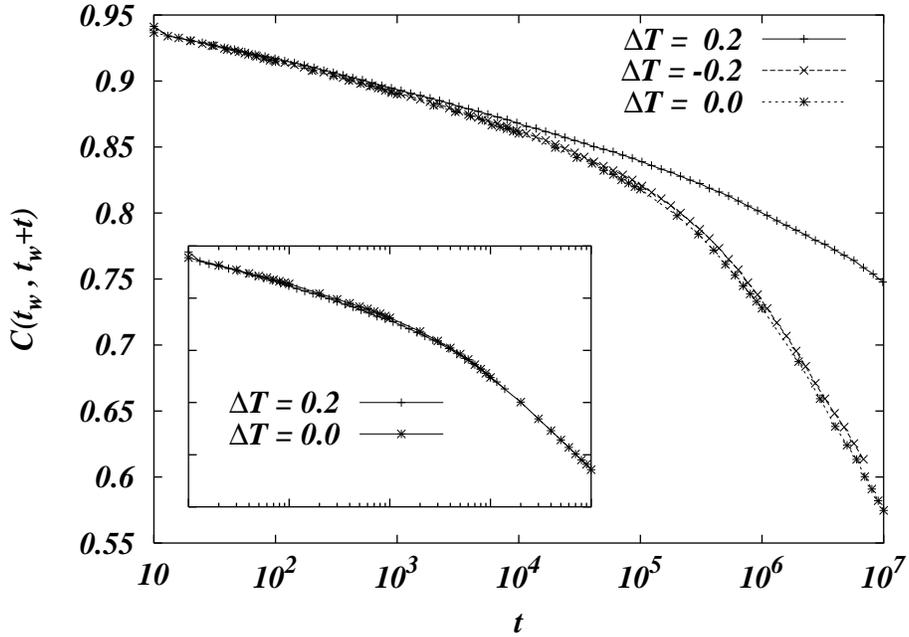}
\caption{Correlation relaxation in temperature cycling experiments.
In the inset we have used temperatures and time scale ratios similar
to the experimental ones (see text for details) and we do not see any
difference with respect to the reference curve. In the main part we
show the results for a large perturbation both in temperature, $\Delta
T = \pm 0.2$, and in times, $t_{w2}=t_{w1}+t_{w3}=10^6$. It is clear
that for both positive and negative cycles the system is more
thermalized with respect to the reference system where $t_w=10^6$.}
\label{corr_ciclo}
\end{center}
\end{figure}

In numerical simulations the effective waiting time (or correlation
time) can be estimated from the decay of the correlation function.  In
the inset of Fig.~\ref{corr_ciclo} we show the correlation measured in
a cycling temperature experiment (with $T_1=0.7$ and $T_2=0.9$) where
the relative times are similar to those used by experimentalists
($t_{w1}=10^4$, $t_{w2}=10^2$ and $t_{w3}=10^2$).  The reference curve
with $\Delta T=0$ will always refer to data measured at fixed
temperature $T_1$ and with a waiting time $t_w=t_{w1}+t_{w3}$.  As it
is clear from the data the temperature cycle does not affect at all
the decay of the correlation function and the system does not seems to
be re-initialized. This effect is still more drastic if we use a
different scheduling in order to amplify it. In the main body of
Fig.~\ref{corr_ciclo} we report the correlations measured with
$t_{w1}= 5 \cdot 10^5$, $t_{w2}=10^6$ and $t_{w3}= 5 \cdot 10^5$, with
$T_1=0.5$ and $T_2=0.7$ or 0.3.  From the data it is clear that the
effective waiting time is increased in both cases in contrast to what
it is observed in experiments, being the time spent at the higher
temperature $T_2=0.7$ much more effective in terms of the
thermalization process.

\begin{figure}
\begin{center}
\epsfxsize=0.7\textwidth \epsffile{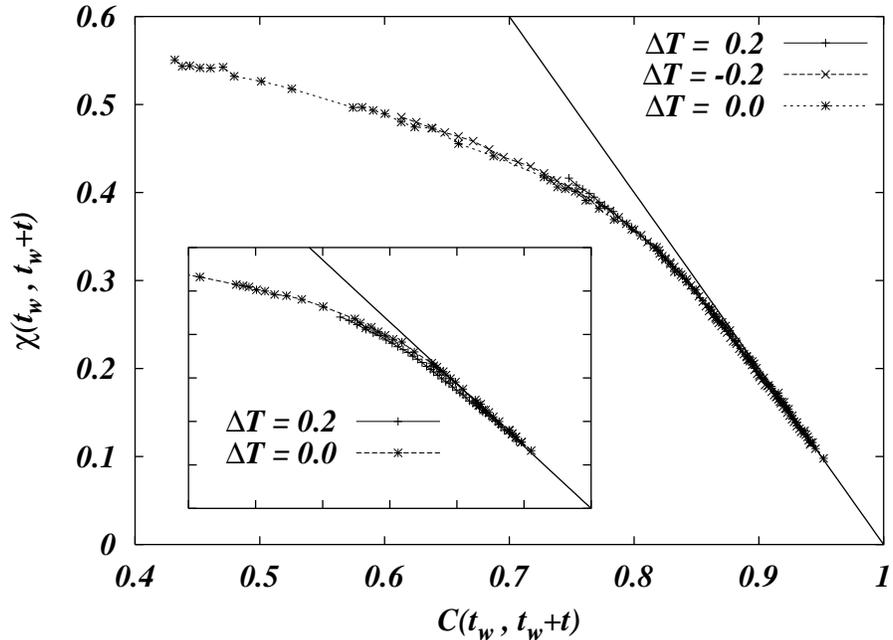}
\caption{Fluctuation-dissipation ratio measured in the experiments of
Fig.~\ref{corr_ciclo}. The (effective) temperature seems to be
unchanged by the temperature cycle.}
\label{fdt_ciclo}
\end{center}
\end{figure}

We expect the effective waiting time to be more or less related to the
size of thermalized regions in the system.  Once we have characterized
how this size changes under a temperature cycle, we can also study how
the internal structure of these thermalized regions is modified by the
cycle.  In order to do this we exploit the generalization of the
fluctuation-dissipation relation to the off-equilibrium regime
discussed in Section~\ref{sec:obs}. In Fig.~\ref{fdt_ciclo} we show
the off-equilibrium susceptibility versus the correlation for the same
experiments reported in Fig.~\ref{corr_ciclo} and described in the
previous paragraph. The temperature cycle, even when it is very long,
does not seem to affect the response of the system. In the
quasi-equilibrium regime the slope of the curve $\chi$ versus $C$
gives the temperature in the thermalized regions.  This temperature
does not seem to change even if the system spends a lot of time in a
different temperature $T_2$. When it comes back to $T_1$ it rapidly
seems to recover the configuration corresponding to temperature $T_1$.

\subsection{Temperature shift experiments}

In order to better understand how the time the system spends at a
temperature $T_1$ can influence the thermalization process at a
different temperature $T_2$ we have performed a series of temperature
shift experiments. Here the scheduling is the following. After
$t_{w1}$ MCS at temperature $T_1$ we set $T=T_2$ and we immediately
start measuring the correlation and the response to a small external
field. In the present study $T_2=0.5$ and $T_1=0.7,0.5,0.3$, the
second case being considered as a reference curve.  Moreover the
waiting times, $t_{w1}=139,10^3,10^5$, have been chosen such that
$t_{w1}^{T_1}$ is constant. It is known that in the Edwards-Anderson
model the dynamical correlation length grows as a power law of time,
$\xi \propto t^{1/z(T)}$, where the dynamical exponent $z(T)$ is
inversely proportional to the temperature~\cite{4DIM,MPRR_00}. So our
choice for the waiting times would correspond to thermalized regions
of similar sizes.

\begin{figure}
\begin{center}
\epsfxsize=0.7\textwidth
\epsffile{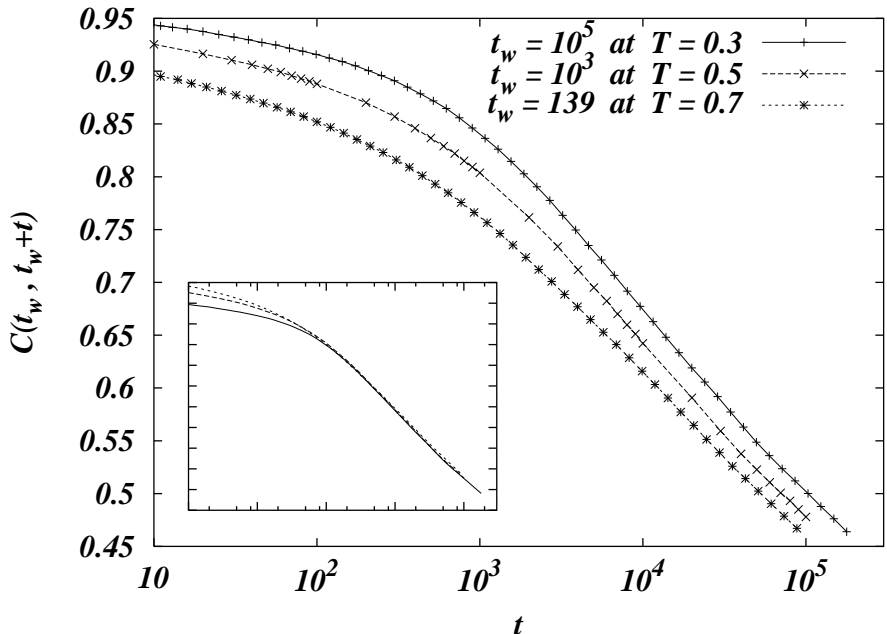}
\caption{The decays of the correlation function at temperature
$T=0.5$, after the thermalizations shown in the legend, have the same
effective waiting time (as can be seen in the inset where we have
rescaled the curves by means of simple multiplicative factors). The
temperatures and the waiting times (see legend) have been chosen such
that the size of thermalized regions, $\xi(t,T) \propto t^{a T}$ (with
$a=0.161$ \cite{MPRR_00} and a prefactor of order 1), is comparable.}
\label{corr}
\end{center}
\end{figure}

In Fig.~\ref{corr} we see that the effective waiting time generated by
the three different scheduling is very similar. The three curves can
be perfectly collapse in the aging regime by simply multiplying them
by a constant (see inset in Fig.~\ref{corr}). This means that the
effective waiting time is essentially given by the size of the
thermalized regions, which has been chosen to be equal in the three
experiments. The difference between the three curves in
Fig.~\ref{corr} comes from the quasi-equilibrium part $C_{st}(t)$,
which decays in a different way.

\begin{figure}
\begin{center}
\epsfxsize=0.7\textwidth
\epsffile{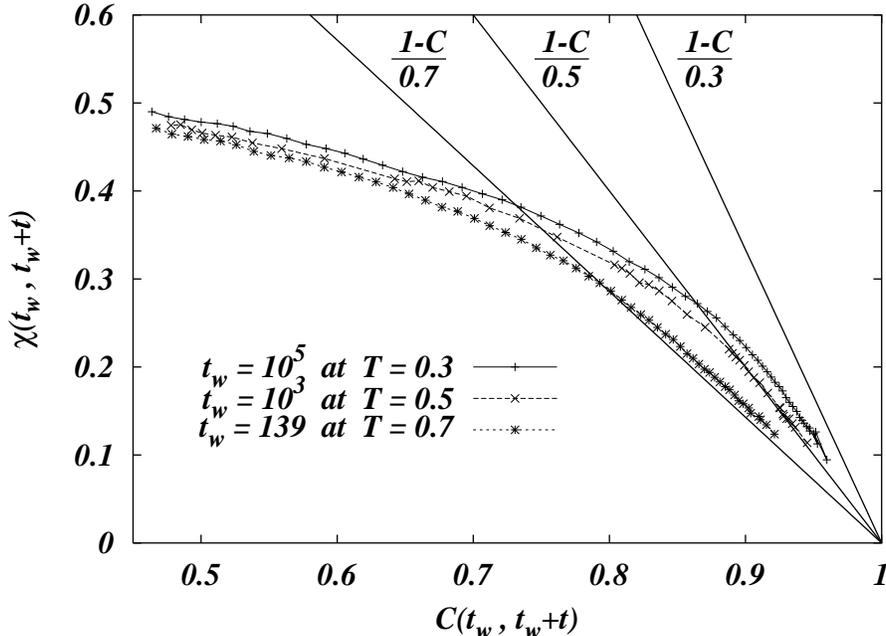}
\caption{FDT measurements taken at a temperature $T=0.5$, after the
thermalizations shown in the legend. They show that in the
quasi-equilibrium regime (i.e.\ in the thermalized regions) the
effective temperature of the system is always the maximum between the
one used to thermalized the system and the one used to take
measurements.}
\label{fdt}
\end{center}
\end{figure}

The following natural question concerns the configuration of the
system up to length scales of the order of $\xi(t)$. Through the
measure of the fluctuation-dissipation ratio we can estimate the
effective temperature of the system on those length scales.  In
Fig.~\ref{fdt} we shown the results for the three experiments. The
reference data, in the quasi-equilibrium regime, perfectly stay on the
line $(1-C)/0.5$ as it should. The other two data sets, because they
have been thermalized at temperature $T_1$ and then let to evolve at a
different temperature $T_2=0.5$, fall in between the line $(1-C)/T_1$
and $(1-C)/0.5$, showing that the system temperature is changing from
$T_1$ to $T_2$. However the interesting point to note is that the
change is very different in the two cases. For $T_1=0.3$ (uppermost
curve in Fig.~\ref{fdt}) the system responds with an effective
temperature very similar to $T_2=0.5$, given by the slope in the
quasi-equilibrium regime. While for $T_1=0.7$ (lowest curve in
Fig.~\ref{fdt}) the effective temperature is perfectly compatible with
$T_1=0.7$.  In general, we would say that if an Edwards-Anderson model
is thermalized to a temperature $T_1$ and measurements are done at a
different temperature $T_2$ the response of the system will be
dominated by the higher temperature. This is the only effect
asymmetric in $\Delta T$ we have found in all the numerical
experiments performed with temperature changes.

Recently Bernardi {\it et al.}~\cite{BERNARDI} have proposed the
following scaling for the susceptibility $\chi(t_w,t_w+t) = \tilde\chi
(\xi(t_w),\xi(t))$, where $\xi(t)$ is the dynamical correlation length
defined above.  Our data for the susceptibility, which are measured on
larger time and temperature scales than ref.~\cite{BERNARDI}, do not
fit that scaling.

\section{Memory and chaos in AC relaxations}
\label{sec:chaos_AC}

In this Section we present a detailed investigation of memory,
rejuvenation and cooling rate effects in the EA model doing AC
susceptibility numerical experiments. In what follows we will use
indistinctly the words chaotic, rejuvenation or reinitialization to
indicate the presence of new relaxational processes which have been
driven off-equilibrium when a perturbation is applied.  There is a
large list of AC experiments in the presence of external perturbations
such as temperature field variations or magnetic field
variations~\cite{SITGES,DC}. The experimental setting is as
follows. The system is quenched to a low temperature (ranging between
$0.6$ and $0.9$ times the value of $T_g$) and the slow decay of the AC
susceptibility recorded. After a time $t_1$ where $\omega t_1$ is such
that the AC susceptibility has not totally decayed to its asymptotic
value a perturbation is applied. To have an idea, the time $t_1$ is
such that both $\chi'$ and $\chi''$ are still between 5\% and 20\%, of
the whole decay, above their asymptotic large $t$ value. This
corresponds to typical values of $\omega t_1$ ranging from $100$ to
$2000$.  At this time there is a sudden perturbation (for instance, a
change in temperature or field). After a time interval $t_2$ which is
of the same order as $t_1$ the perturbation is switched off. In the
present study and for sake of simplicity we have taken $t_2=t_1$.  
All the times $t$ and frequencies $\omega$ we used in the
numerical experiments are such that the scaling $\omega t$ is
satisfied~\cite{SCAL_CORR}.

In the presence of rejuvenation or chaotic effects one generally
observes strong reinitialization of the AC susceptibilities
corresponding to processes which have been driven off-equilibrium as
consequence of the perturbation. Having in mind the previous
experimental setting we have considered the following different types
of perturbations: temperature changes, magnetic field changes and
quenched disorder changes.

\subsection{Temperature changes}

\begin{figure}
\begin{center}
\epsfxsize=9cm
\epsffile{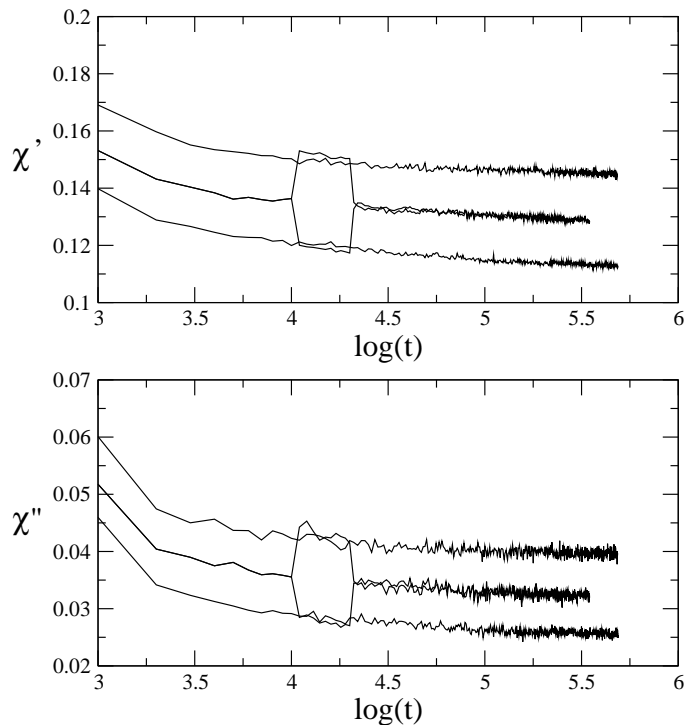}
\caption{AC temperature variation experiment. The system is quenched
at $T=0.6$ and the AC susceptibility is recorded for field of period
$P=100$ and intensity $h_0=0.1$. At time $t_1=10000$ the temperature
is changed by $\Delta T=\pm 0.1$ and after the same time interval
temperature is restored to its original value.  As comparison we show
the reference relaxation curves at temperatures $T+\Delta T=0.7$ and
$T-\Delta T=0.5$}
\label{64T06P100}
\end{center}
\end{figure}

\begin{figure}
\begin{center}
\epsfxsize=9cm
\epsffile{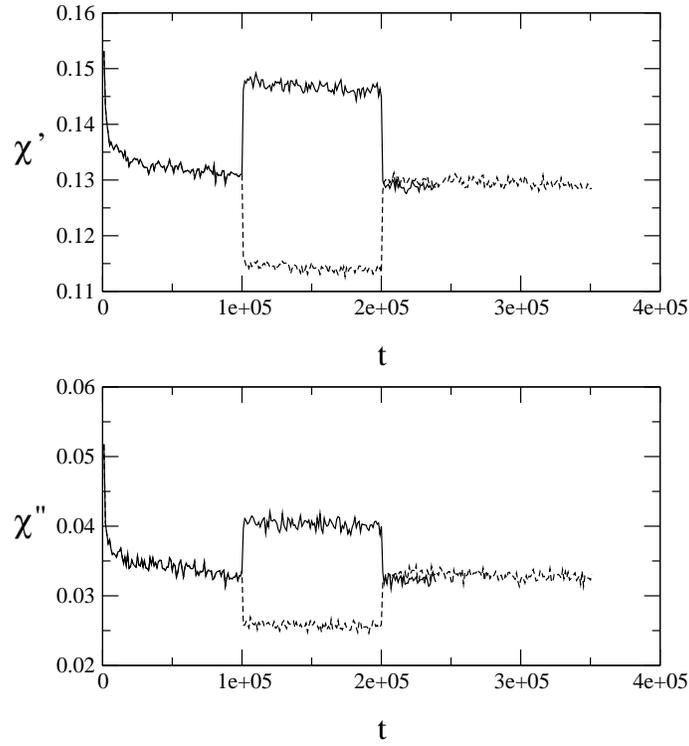}
\caption{The same experiment as in Fig.~\ref{64T06P100} but with
$t_1=100000$.}
\label{64T06P100_2}
\end{center}
\end{figure}

We quench the system to a temperature below $T_c$ and apply an AC
magnetic field measuring the AC susceptibility. At a given time $t_1$
we suddenly change the temperature to $T+\Delta T$ and still measure
the AC susceptibility. Then, after a time interval equal to $t_1$ we
reset again the temperature to its original value $T$. In the presence
of chaotic or rejuvenation effects we expect that a sudden change in
temperature will reinitialize some relaxational processes. In
Fig.~\ref{64T06P100} we show the results for $L=64$ by measuring
relaxation at $T=0.6$ and making two jumps in temperature $\Delta
T=\pm 0.1$ at times $t_1=10000$ and $t_1+t_2=2t_1$. The jump in
temperature is then applied when a large part of the AC susceptibility
is still relaxing like in the experimental setting. Note that for a
positive temperature change $\Delta T$ the AC susceptibility stays
above the reference curve at the temperature $T+\Delta T$. This means
that the effective waiting time after the positive jump is smaller
than that of the reference curve at higher temperature.  For a
negative temperature jump $-\Delta T$ the AC susceptibility stays
below the reference curve at temperature $T-\Delta T$. This means that
the effective waiting time has now increased and relaxation at
$T-\Delta T$ has benefited from relaxation at the higher temperature
$T$.  The results of Fig.~\ref{64T06P100} show that the effective time
$t_{\rm eff}$ during the interval of time $t_2$ when temperature has been
changed is controlled by the same activated processes but with a
different activation rate,
\begin{equation}
t_{\rm eff}=t_2^{\frac{T\pm \Delta T}{T}} \quad ,
\label{effective}
\end{equation}
implying $t_{\rm eff}> t_2$ if $\Delta T >0$ and vice versa.  As
comparison we also show a similar plot for $t_1=100000$ when nearly
all relaxation of AC susceptibility has taken place in
Fig.~\ref{64T06P100_2}. Again no traces of re-initialization effects
in the AC susceptibility are observed.  The main feature we can
appreciate from the figure is that the positive jump and the time
spent at $T+\Delta T$ has increased the effective age of the system
respect to the reference temperature $T$ respect to the curve for the
negative jump in agreement with Eq.(\ref{effective}). Note in the
figure that, after the second jump, the dashed line stays above the
continuous line. Note that this temperature dependence of the
effective time (\ref{effective}) is the one found in the
REM~\cite{REM,SN1} and here seems to behave quite well.

\begin{figure}
\begin{center}
\epsfxsize=9cm
\epsffile{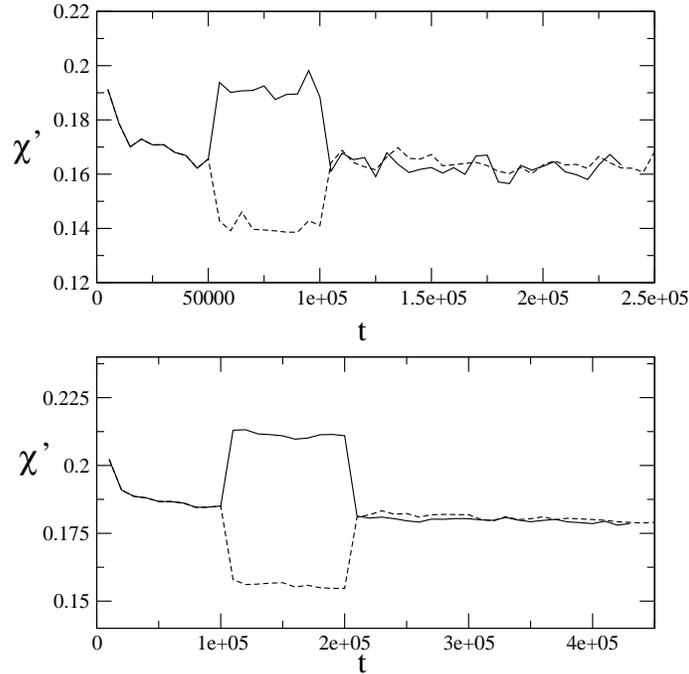}
\caption{Plot above: The equivalent experiment of Fig.~\ref{64T06P100}
but with a larger size, a smaller field and a smaller frequency:
$L=100, h=0.01, P=500$. Plot below: The equivalent experiment of
Fig.~\ref{64T06P100} but with a 10 times smaller frequency
$P=1000$. Note that the perturbation in these two experiments and in
Fig.~\ref{64T06P100} are applied when $\omega t=100$.}
\label{100T06P500h001}
\end{center}
\end{figure}

The picture which emerges from these figures is in agreement with all
the results published up to now which point in the direction that
there are no chaotic or rejuvenation effects in Edwards-Anderson spin
glasses in the presence of temperature changes. One could argue that
there are several factors which induce the absence of {\em any trace}
of rejuvenation. Among them: 1) The intensity of the field which is
bigger than in experiments; 2) The smallness of the period of the
oscillating field which covers at most {\em nanoseconds} when compared
to real experiments (of order of seconds) and 3) The size of the
system which is large enough.  In Fig.~\ref{100T06P500h001} we show
the same results as Fig.~\ref{64T06P100} for a field $h_0=0.01$, a
larger size $L=100$ and a larger period $P=500$ such that $\omega t$
takes the same value, so we are in the same time scale according to
the results of ref.~\cite{SCAL_CORR}. Due to the smallness of the
probing field the signal is now much more noisy so we show the
in-phase susceptibility.  The smallness of the absolute magnitude of
the time scales involved in numerical experiments (reason 2) above) is
usually advocated as the main source of discrepancy between numerics
and experiments. The lower plot in Fig.~\ref{100T06P500h001} shows
$\chi'$ for the same size and field as Fig.~\ref{64T06P100} but with a
frequency 10 times smaller.  The conclusions are exactly the same:
cooling rate effects are important and no trace of reinitialization
after the temperature variation is observed.

\subsection{Field and coupling distribution variations}

To make evident how much this absence of chaos or rejuvenation is
indeed an intrinsic effect to temperature changes we have done the
same experiment with a different type of perturbation. Instead of
changing the temperature, we have applied a perturbation which is well
known to be chaotic from equilibrium studies.  Examples of such
perturbations are: 1) a change in the uniform magnetic
field~\cite{KONDOR,RITORT} and 2) a change in the couplings
distribution~\cite{AFR}.

\begin{figure}
\begin{center}
\epsfxsize=9cm
\epsffile{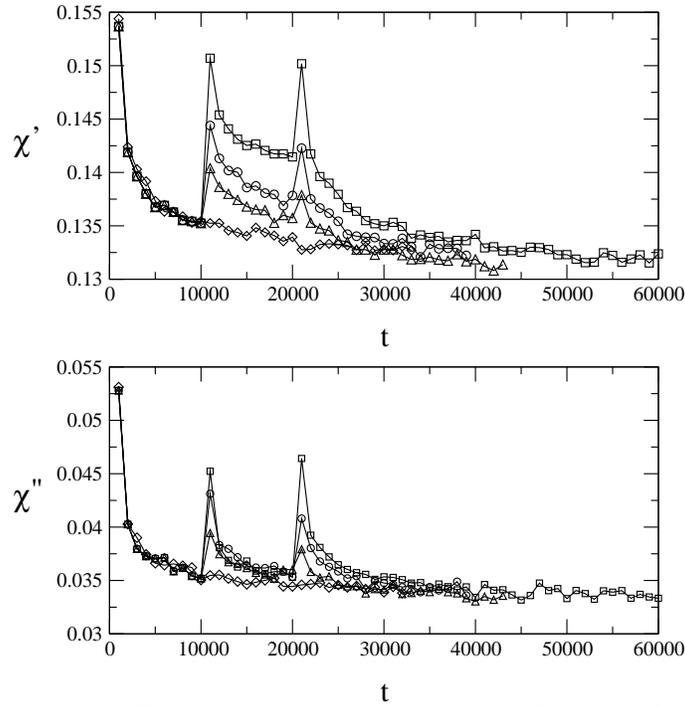}
\caption{DC Field variation experiment. The system $L=100$ is quenched
at $T=0.6$ and the AC susceptibility is recorded for a probing field
of period $P=100$ and intensity $h_0=0.1$. At time $t_1=10000$, a DC
field is applied for a time interval $t_2=t_1$. After $t_1+t_2$ the
field is switched off. The intensities of the DC fields are
$h_{DC}$=0.4 (triangles), 0.6 (circles) and 1.0 (squares). The
diamonds correspond to the reference curve without perturbation.}
\label{hchi}
\end{center}
\end{figure}

Concerning the first type of variation there have been several
experiments which reveal how reinitialization occurs under a DC
magnetic field change~\cite{DC}. The experimental setting is the same
as that shown previously but now the perturbation is to apply a DC
magnetic field after $t_1$. So the system is quenched at zero DC field
and at time $t_1$ the DC field is switched on. After a time interval
$t_2=t_1$ the DC field is set to zero again. In laboratory
experiments~\cite{DC} the intensity of the applied DC field must be
larger than the amplitude of the AC field for the AC field to probe
the response of the system after the DC perturbation. The intensity of
the probing AC field is typically smaller than one Oersted and the
intensity of the AC field much higher (between 5 and 10 Oersteds). So,
typically the intensity of the DC field is 10 times or even more
larger than the probing AC field. Nevertheless, in the numerical
experiments we have a problem. The intensity of the AC field cannot be
arbitrarily small, otherwise we have a too small and noisy
signal. Consequently, if the perturbing DC field is chosen 10 times
larger than the AC field then the resultant field (AC+DC) will be very
large and drive the system out of the linear response
regime. Moreover, non-linear effects will be much enhanced because of
the non-linear coupling between the perturbation (the DC field) and
the probing field (the AC field). A way to avoid this is to apply a
perturbing DC field which does not couple with the probing field, for
instance, a staggered DC field. Now the perturbation is given by a new
term $\delta{\cal H}$ in the Hamiltonian
\begin{equation}
\delta{\cal H}=-h_{DC}\sum_{i=1}^V\eps_i\s_i \quad ,
\label{stag}
\end{equation}
where $\eps_i$ are quenched random variables which may take the values
$\pm 1$. Another equivalent procedure would be to apply a
uniform DC field as perturbation and measuring the AC susceptibility
corresponding to the response to a staggered probing AC field. The
results are shown in Fig.~\ref{hchi}. Note that strong
re-initialization is seen after perturbing the system in agreement with
the known result that finite-dimensional spin glasses are chaotic
against magnetic field changes~\cite{RITORT}.

\begin{figure}
\begin{center}
\epsfxsize=9cm
\epsffile{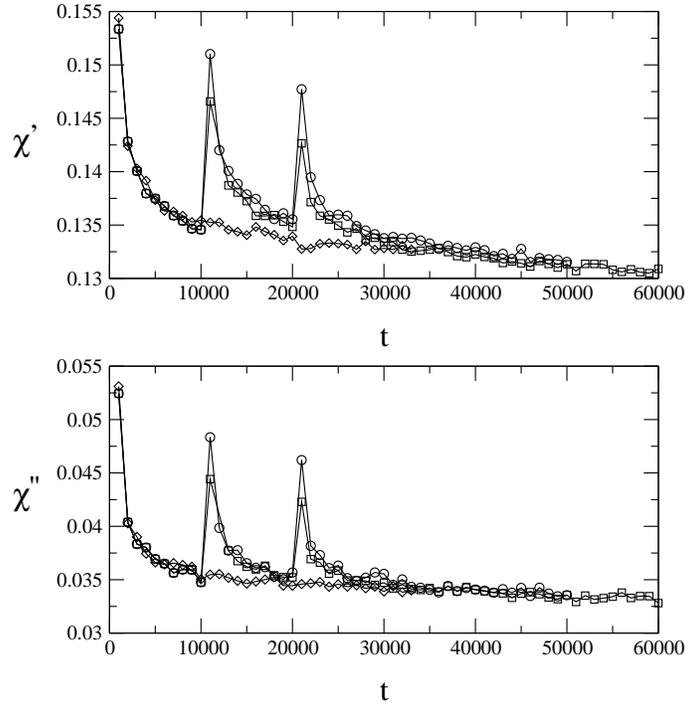}
\caption{Couplings variation experiment. The system $L=100$ is
quenched at $T=0.6$ and the AC susceptibility is recorded for a
probing field of period $P=100$ and intensity $h_0=0.1$. At time
$t_1=10000$, a percentage $r$ of the couplings change sign and the AC
susceptibility is recorded for a time interval $t_2=t_1$. After
$t_1+t_2$ the couplings take again their original values. The
intensities of the perturbation are $r=0.05$ (squares) and $r=0.1$
(circles). The diamonds correspond to the reference curve without
perturbation.}
\label{rchi}
\end{center}
\end{figure}

A similar result is found by considering the other type of
perturbation.  In that case we measure the AC susceptibility after
quenching at temperature $T$. At $t_1$ we take a percentage $r$
($0<r<1$) of the couplings and reverse its sign
$J_{ij}=-J_{ij}$. After a new interval $t_1$ we re-put the original
couplings again. The results are shown in Fig.~\ref{rchi} for
$t_1=10000$ $\omega=0.01$ and $r=0.05,0.1$ (corresponding to 5\% and
10\% of changes in the couplings respectively). Note the presence of
strong and clear re-initialization effects in agreement with the fact
that such a perturbation is chaotic.

We may conclude this Section saying that while there is clear trace of
chaotic behavior in the presence of field or couplings changes, there
is absolutely no trace of re-initialization effects below the
spin-glass transition in the case of temperature change experiments.
This may be due to the presence of cooling rate effects in three
dimensional Ising spin glasses stronger than those measured in
laboratory experiments.

\section{Outlook and discussion}

Two years ago some experimentalists from the Saclay and the Uppsala
groups measured on a spin glass sample very strong memory and chaos
effects~\cite{MEMORY_EXP}. Their results are really impressive and
show unambiguously how important are these effects in real spin
glasses.  By investigating the Edwards-Anderson model in three
dimensions we have tried to reproduce numerically their findings, but
we have obtained results pointing in the opposite
direction. Temperature variation experiments in spin glasses are
nowadays one of the most puzzling results in the field. It is unclear
which is the final theory which may naturally account for these
results. It is not easy to explain, from the point of view of the
droplet model~\cite{DROPLET}, how reorganization of domains can
account for the memory effects observed in the
experiments~\cite{MEMORY_EXP,RECENT_SG,BOUCHAUD}.  Already in
equilibrium theory, the question whether the spin-glass phase is
chaotic against temperature changes is far from being settled.
Extensive numerical work does not show any clear evidence of chaos for
temperature changes~\cite{RITORT,KVFN,NY,MB}.

Regarding off-equilibrium dynamics the situation is similar. Very
precise numerical simulations by Rieger et al.~\cite{KSSR} show that
correlations between equilibrium configurations at different
temperatures are big and the corresponding overlap length grows
algebraically in time without any tendency to saturation within the
simulated range of times.  The same conclusions hold for TRM numerical
experiments with temperature change protocols~\cite{RIEGER}. Very
recently, Komori et al.~\cite{KYT} have presented a detailed study of
the two-time correlations in the presence of temperature change
variations. These should be essentially equivalent to the present
results because AC experiments probe the quasi-stationary aging regime
where fluctuation-dissipation makes responses and correlations
equivalent. Their conclusion is like ours: no rejuvenation effects are
found and cooling rate effects are very strong.  The only claimed
evidence for rejuvenation and memory effects is the Figure 7 in that
reference which is unconclusive as we have explained
in Section~\ref{sec:chaos_TRM}. Contrarily, the results for DC field
variation experiments resemble quite much those of experiments and,
together with coupling distribution variations, show that chaotic
effects manifest as re-initialization effects in the AC
susceptibility.

What is the origin of this discrepancy between experiments and
simulations? We indicate three possible reasons:

\begin{enumerate}

\item The size of the system is too small.

\item The intensity of the probing field is too large.

\item The absolute magnitude of the time scales in numerical
experiments are too small.

\end{enumerate}

As we showed in Fig.~\ref{100T06P500h001} improvement in these
limitations (larger size, smaller frequency, smaller AC field) does
not seem to alter the final conclusions. Unfortunately we cannot still
definitely conclude anything because probably our improvements (within
the available computer capabilities) are too modest. Let us briefly
comment about these three possibilities. Probably the finiteness of
the size is the less important. A larger size diminishes statistical
fluctuations but has no strong effect on the growing domains because
they are known to be very small~\cite{HUSE,KSSR,MPRR,KYT2}.  Regarding
the smallness of the field we have shown in Fig.~\ref{100T06P500h001}
that a field 10 times smaller does not change the conclusions and for
this smaller value of the probing AC field we are much closer in
magnitude to the experimental setup. Moreover, field-cooling and
zero-field cooling experiments, using the same values for the external
DC field, show that the system is in the linear response
regime~\cite{MPRR,PRR_99}. There are no deep reasons why things should
drastically change for an AC probing field 100 times smaller to the
one used in the present simulations. At most the intensity of the
field can give non negligible corrections to the usual $t/t_w$
scaling~\cite{PARISI} as shown in~\cite{SCAL_CORR}.

The smallness of time scales involved (i.e.\ the fact that AC
frequencies are too large and the waiting times too small) is the most
serious reason and could definitively be the origin of the
discrepancy. Short-time scales obviously imply short length scales of
size $\xi(P)$ or $\xi(t_w)$ depending on the kind of experiment.  If
temperature changes would have no effect on small length scales but
only on very large scales then chaotic effects could not be seen in
standard numerical simulations.  This statement corresponds to saying
that the overlap length $L(\Delta T)$ for a typical temperature change
$\Delta T$ is much larger than any probed domain length in the
numerical experiment, but smaller than those probed in the laboratory
(where the chaos is clear).  Let's try to quantify more this
statement.  In laboratory experiments~\cite{XI_EXP} typically length
scales of the order of $\xi \sim 10^2$ can be reached, while in
numerical simulations we are restricted to length scales between $3$
and $5$~\cite{KSSR,MPRR_00}.  This discrepancy may or may not be a
deep trouble depending on the value of the chaos
exponent. Unfortunately there is not a numerical estimate for the
chaos exponent (because chaos in temperature has never been
observed!!!) and we only have an estimate from domain-wall scaling
arguments~\cite{BM} which give $L(\Delta T)=(\Delta T)^{-\zeta}$ with
$\zeta=d_s/2-\theta$ where $d_s$ is the fractal surface of droplets
and $\theta$ is the thermal exponent. For $d=3$, the value of $\zeta$
must be larger than $1$ and a reasonable value seems to be
$\zeta\simeq 1.5$~\cite{MH} so chaos in temperature should not be too
small after all. Note also that for a DC field change the chaos
exponent is even smaller, the overlap length being given by $L(h)\sim
h^{-\frac{2}{3}}$,~\cite{RITORT}. So, for similar values for the
prefactor, one would expect stronger chaos in temperature than in a DC
field. On top of that, the analysis by Bray and Moore~\cite{BM} for
the value of the chaos exponent $\zeta$, shows that is consequence of
the balance between a contribution coming from the surface of the
droplets $L^{\frac{d_s}{2}}$ and a contribution from the activation
energy necessary to revert a droplet $L^{\theta}$. The surface
contribution comes from the inhomogeneities in the couplings on the
surface and, actually, the same type of analysis should fully carry
through when analyzing chaos for coupling perturbations. But in this
last case we found strong re-initialization effects (see
Fig.~\ref{rchi}) which are absent when changing the temperature. How
this difference which we observe can be explained in the framework of
the droplet model remains mysterious unless one advocates different
prefactors for both overlap lengths (much bigger for temperature
changes than for couplings changes) or that something special occurs
in three dimensions~\cite{HN}. With the present estimations for the
$\zeta$ exponent and if the prefactor for the overlap length is very
large, chaotic effects, which experimentally appear on temperature
changes such that $\Delta T \simeq 0.1~T_c$, may need in present
simulations temperature changes of the same order of the absolute
temperature. If this is the case then we have to wait for the next
computer generation or to study the EA model in a situation such that
larger length scales could be reached (e.g.\ in 4d or in 3d
with next-nearest neighbours interactions).

On the other hand our results imply that if the laboratory experiments
were done at frequencies of $10^9$ Hertz (instead of the typical 1 Hz
measurements) then cooling rate effects would be restored and
rejuvenation or chaotic effects disappear. Unfortunately, there are no
experiments in the range $10-10^9$ Hertz. Note also that, according to
the general $\omega t$ scaling, for these frequencies one should do
measurements at very short times such that $\omega t$ is not much
larger than $10^4$ and the AC susceptibility has not completely
relaxed. Still, the results of ref.~\cite{SCAL_CORR}, the experimental
results of~\cite{SITGES} and all the numerical published data up to
now show that the scaling $\omega t$ (the equivalent of the scaling
$t/t_w$ in two-time experiments) works reasonably well (with some
slight deviations). If the scaling $t/t_w$ means something (as most of
the present theoretical work suggests) then it is difficult to
understand why no trace of re-initialization effects is observed in
the smallest coarsening domains when the temperature is
changed. Actually these re-initialization effects are found for DC
magnetic field and coupling distribution changes. If some dynamical
effects are completely absent for the small coarsening sizes this
means that no numerical simulation in the last ten years has actually
reached the asymptotic regime where connection to real experiments is
possible and we are certainly missing something~\cite{BOUCHAUD}.

Concerning theory, it is difficult to give a complete description on
these effects in terms of compact excitations as proposed in the
droplet model~\cite{DROPLET}. Usually the asymmetric response of the
spin glass against the sign of the temperature perturbation is
explained in terms of an asymmetric overlap length $L(\Delta T)$. The
problem is that in simulations $L(\Delta T)$ is apparently extremely
large when compared to experiments and also symmetric, a question
which is difficult to explain again if one does not appeal to the
smallness of the time scales involved. If the necessary time scales to
see the asymmetric effects in $L(\Delta T)$ {\em exceed} the
experimental time scales~\cite{BDM} then it is unclear how to
reconcile any two among the three: theory, experiments and
simulations.

The explanation of memory and chaotic effects was originally explained
in terms of a hierarchical picture~\cite{SITGES}. Actually, these
effects are explained and reproducible in the GREM~\cite{GREM,SN2}, a
model with several critical temperatures. The same phenomena is absent
in the random-energy model~\cite{REM,SN1} with a single critical
temperature. Unfortunately, the GREM is a model without microscopic and
spatial description so the connection with real spin glasses remains
speculative. As we said previously the results we find here, and in
particular the cooling rate effects, are very similar to those of the
random energy model~\cite{SN1}. It is well known that the spin-glass
phase in this model can be described in terms of a one-step
solution. And what we see in our simulations is what is expected for a
mean-field model with a one-step of replica symmetry breaking
solution~\cite{CK}. This solution is known to describe the physics
behind structural glasses~\cite{BCKM}. Actually temperature variation
experiments on structural glasses~\cite{LN,BCL,ADL} resemble our
simulation results much more than what experiments on spin glasses
do. Again one could claim that the one-step character of the effects we
observe in our simulations are due to the smallness of time scales and
if one increases the time scales (let us say by 6 orders of magnitude)
much different results will come out.

But there could be another and more natural explanation. Is the
Edwards-Anderson model really a good spin glass? Is it possible that the
Edwards-Anderson model has strong cooling rate effects and no
rejuvenation effects at all? If rejuvenation or chaotic effects are
present in the dynamics for a given perturbation this could be related
to the fact that equilibrium properties are chaotic when such a
perturbation is switched on.  Analytical calculations in the mean-field
version of the Edwards-Anderson model (the Sherrington-Kirkpatrick
model) confirming chaoticity for temperature changes are still
inconclusive~\cite{MB,RIZZO}. A completely different scenario holds for
field and coupling perturbations in agreement with the present
simulations.  Let us note also that real spin glasses are site
disordered systems, a type of disorder not included in the
Edwards-Anderson model. Recent experiments on the Kagome antiferromagnet
lattice~\cite{WDVHC} reveal that there are not strong re-initialization
effects after changing the temperature. Because that system does not
include disorder at all one may wonder whether the influence of site
disorder can be important. Site-disordered spin glasses were not much
considered in the past (for some works and references see~\cite{SITE})
because they were thought to be less relevant to experiments than
bond-disordered models but probably this is not true and a site disorder
effect should be taken into account to explain experiments. To finish
this collection of possible ways out to this puzzle let us point out
also the possible role of the continuous character of the spins in a
real spin glass as well as the effect of chirality \cite{CHIRAL}. This last effect and
its importance in the description of the dynamics of spin glasses will
surely see further developments in the forthcoming years.

It is clear that we are facing a very difficult problem. If the
smallness of time scales of the simulation is the final explanation
for everything then a sensible theory for spin glasses must explain
why temperature changes are fundamentally so peculiar when compared to
other type of perturbations, i.e. why the prefactor for the overlap
length $L(\Delta T)$ is so large. In this respect, experiments at much
larger frequencies are necessary. At least, to see if cooling rate
effects gradually change and re-initialization effects, in the presence
of temperature variations, do systematically weaken. On the other
hand, if the Edwards-Anderson model fails to explain a crucial result
found in experiments then we must discover what ingredient is lacking
in the original model and what are the consequences for our present
knowledge of spin glass theory. In this direction, finding a
microscopic model with spatial structure which presents a spin-glass
phase transition and unambiguously shows memory, chaotic as well as absence of
cooling rate effects would be welcome.

{\bf Acknowledgements} We are grateful to J.P. Bouchaud, M. Mezard,
G. Parisi and E. Vincent for useful discussions and H. Rieger and
H. Takayama for useful correspondence.  M.P. and F.R. acknowledge
financial support from a French-Spanish collaboration (Picasso program
and Acciones Integradas Ref. HF1998-0097).

\end{document}